\documentclass[journal=jacsat,manuscript=article,layout=twocolumn]{achemso}

\usepackage[version=3]{mhchem} %
\usepackage{graphicx}%
\usepackage{dcolumn}%
\usepackage{bm}%
\usepackage{subcaption}
\usepackage{arydshln}
\usepackage{tabularray}
\usepackage[utf8]{inputenc}
\usepackage[T1]{fontenc}
\usepackage{mathptmx}
\usepackage{etoolbox}
\usepackage{makecell}
\usepackage[dvipsnames]{xcolor}
\usepackage{titlesec}

\titleformat*{\section}{\large\bfseries}
\titleformat*{\subsection}{\normalsize\bfseries}
\titleformat*{\subsubsection}{\normalsize\bfseries}

\author{Alexander Chen}
\affiliation[RPI]{Department of Physics, Applied Physics and Astronomy, Rensselaer Polytechnic Institute, Troy, New York}

\author{Meng Zhang}
\affiliation[RPI2]{Department of Electrical, Computer, and Systems Engineering, Rensselaer Polytechnic Institute, Troy, New York}

\author{Daniel Crowley}
\affiliation[RPI2]{Department of Electrical, Computer, and Systems Engineering, Rensselaer Polytechnic Institute, Troy, New York}

\author{Nicholas Gangi}
\affiliation[RPI2]{Department of Electrical, Computer, and Systems Engineering, Rensselaer Polytechnic Institute, Troy, New York}

\author{Amir Begovi\'{c}}
\affiliation[RPI2]{Department of Electrical, Computer, and Systems Engineering, Rensselaer Polytechnic Institute, Troy, New York}

\author{Z. Rena Huang}
\email{*huangz3@rpi.edu}
\affiliation[RPI]{Department of Physics, Applied Physics and Astronomy, Rensselaer Polytechnic Institute, Troy, New York}
\alsoaffiliation[RPI2]{Department of Electrical, Computer, and Systems Engineering, Rensselaer Polytechnic Institute, Troy, New York}

\title%
  {\fontfamily{ptm}\selectfont Silicon Photonics Foundry Fabricated, Slow-Light Enhanced, Low Power Thermal Phase Shifter}

\abbreviations{IR,NMR,UV}

\begin{document}
\begin{tocentry}
\includegraphics[width=8.25cm, height=4.45cm, keepaspectratio]{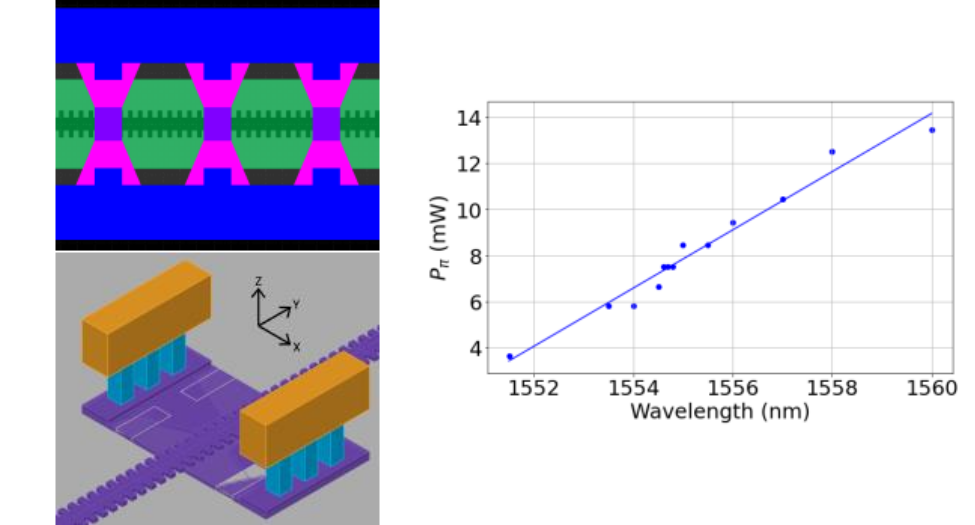}

\end{tocentry}

\begin{abstract}
 
 In this research, we developed a low-power silicon photonics foundry-fabricated slow-light thermal phase shifter (SLTPS) where the slow-light (SL) effect is achieved using an integrated Bragg grating (BG) waveguide. Heating the grating induces a red shift in the transmission spectrum, leading to an increased group index \(n_g\) during operation, which facilitates a further reduction in the voltage needed for a \(\pi\) phase shift, i.e. \(V_{\pi}\). Additionally, we investigated a compact Mach-Zehnder Interferometer (MZI) that incorporates the SLTPS in both arms with a phase shifter length of 50 $\mu$m. A detailed theoretical analysis was conducted to address the non-idealities of the SL-MZI due to uneven optical power splitting and unbalanced loss in the two MZI arms. The \(V_{\pi}\) and power consumption for a \(\pi\) phase shift (\(P_{\pi}\)) of the SL-MZI were quantified for operation in the slow light regime, demonstrating a \(V_{\pi}\) of 1.1 V and a \(P_{\pi}\) of 3.63 mW at an operational wavelength near the photonic band edge. The figure of merit (FOM) \(P_{\pi} \times \tau\) is commonly used to assess the performance of thermal optical switches. The SL-MZI in this work has achieved a low \(P_{\pi} \times \tau\) of 5.1 mW \(\mu\)s. Insertion loss of the SL-MZI ranges from 1.3 dB to 4.4 dB depending on the operation wavelength, indicating a trade-off with the $V_\pi$ reduction.

\end{abstract}

\section*{Introduction}
\label{sec:Introduction}

In recent years, silicon photonics has emerged as a promising platform for high-performance, scalable integrated photonic systems. Crystalline silicon (Si), with a thermo-optic coefficient of $ 1.84\times10^{-4} \text{K}^{-1}$, \cite{espinola2003fast} is well suited for constructing thermally tunable functional devices such as optical switches\cite{coppola2011advance}, tunable delay lines\cite{hailu2019tunable}, filters, and more.  The thermo-optic phase shifter (PS) serves as a fundamental building block for numerous complex integrated lightwave circuits. Consequently, thermal PS on Si platform has been developed and is now readily accessible through the processing development kit (PDK) libraries offered by many Si photonics (SiPho) foundries \cite{jacques2019optimization}. The availability of a standard foundry processing-compatible Si thermal PS with low power consumption and a compact form factor, is of immense significance and plays a paramount role in the scalability of photonic systems across a wide range of applications, including optical beamforming in phased array antennas \cite{zhou2017developing}, chip-scale photonic accelerators \cite{clements2016optimal, shen2017deep, harris2018linear, zhou2022photonic,zhang2024tempo}, and the realization of quantum gates for on-chip photonic computing \cite{zhang2021supercompact}.

Various thermal strategies have been explored including heating element placed above the waveguide \cite{sun2010submilliwat,schall2016infrared,yan2017slow,parra2020atomic}, in parallel \cite{vancampenhout2010integrated,patel2014four,yan2018efficient}, or embedded within the waveguide \cite{watts2013adiabatic,harris2014efficient}. Different materials such as metallic electrodes \cite{sun2010submilliwat}, doped silicon \cite{patel2014four}, and graphene \cite{schall2016infrared,yan2017slow} have been implemented to generate heat via current conduction. The performance of Si thermal optical switch is quantified by the FOM, $P_\pi\times\tau$, which assesses the heating efficiency and switching speed. Here, $P_\pi$ is the power required to achieve $\pi$ phase shift and $\tau$ is the time constant of the thermal PS. A common trade-off between $P_\pi$ and $\tau_\pi$ is often observed in thermal PS design \cite{watts2013adiabatic,harris2014efficient}. At present, Si thermal PS using foundry compatible process can achieve FOM $\sim$50 mW$\mu$s \cite{patel2014four,harris2014efficient}. Graphene microheaters on photonic crystal (PhC) waveguides  \cite{schall2016infrared,yan2017slow} have demonstrated a remarkable FOM of 2.37 mW$\mu$s; however 2D material graphene integration isn't compatible with SiPho foundry process at present.  Thermal PS on free-standing Si \cite{sun2010submilliwat} can achieve sub-milliwatts switching power with minimal thermal crosstalk, but has a very long switch time and also requires substantial modifications to the existing SiPho foundry process \cite{yan2018efficient}.

Thermal-optical PS operating in slow-light mode has emerged as a foundry-compatible solution, offering significantly reduced switching voltage $V_\pi$ and power. Previous research has explored single-line defect PhC waveguide, engineering the dispersion to achieve a constant group index ($n_g$) of $\sim$30 \cite{yan2018efficient,yan2017slow} . In this study, we explore a simpler device structure using a one dimensional dielectric PhC, namely a Bragg grating (BG) to enhance FOM. Rather than using a constant $n_g$, the operation of the BG based thermal PS induces a red shift of the band edge, leading to a monotonic  increasing of $n_g$ during off-to-on switching transient that can greatly reduce $V_\pi$ and $\tau$. Typically, the fall time $t_f$ of thermal PS dominates the total switching delay. Our experimental testing reveals that the lowest temperature increase required to achieve a $\pi$ phase change $\Delta T_\pi$  is 18.4$^\circ$C. By reducing the raising temperature, we can significantly decrease the thermal PS heat dissipation time $t_f$.  In this work, the thermal PS has achieved a FOM of 5.1 mW$\mu$s experimentally, while the devices were fabricated at the American Institute of Manufacturing Photonics (AIM Photonics) AIM foundry without any process modifications, making it readily adoptable in complex photonic system realization.

\section*{Device Design and Fabrication}

\begin{figure}[h!bt]
    \centering
    \includegraphics[width=.95\columnwidth]{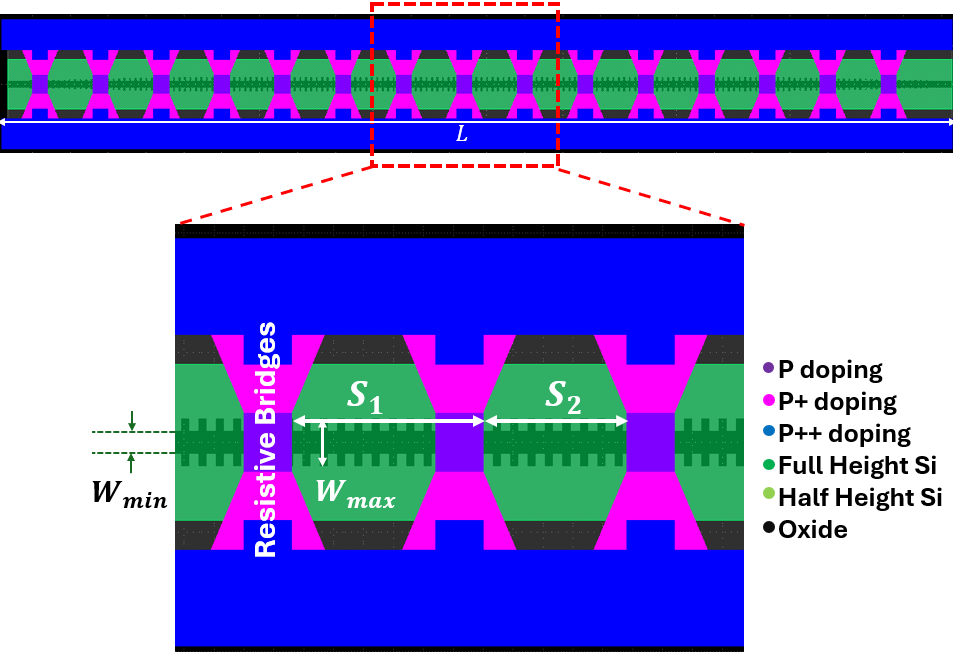}
    \vspace{-10pt}
    \caption{The $L=$50 $\mu$m SLTPS grating with a close-up view of the boxed grating section. P-doped regions of the SLTPS. Doping comes in three concentrations: P, P+ and P++. The distance between the grating structure and the P++ type area is 1.4 $\mu$m. The separation distance between each resistive bridge is $S_2=$2.4 $\mu$m which is on a period of $S_1$=3.2 $\mu$m. }
    \label{fig:grating2}
\end{figure}

The SLTPS device discussed in this study was fabricated at the AIM foundry as part of a multi-project wafer (MPW) active run. The BG thermal PS is placed in both arms of a MZI configuration. Those two arms have a center-to-center distance of 25 $\mu$m. The MZI comprises two pairs of directional couplers for beam splitting and combining. The layout of the SLTPS is illustrated in Fig.~\ref{fig:grating2}. The AIM PDK library provides a thermal PS component with a PS length of 100 $\mu$m, characterized by a $V_\pi$ of 3 V and $P_\pi$ of 30 mW. In pursuit of a more compact device, in this work, the SLTPS length is reduced to 50 $\mu$m in the design. P-doped Si bridges, positioned with a period of $S_1=$3.2 $\mu$m along the Si ridge waveguide, were introduced to create an embedded heating structure. Each heater bridge has a width of 800 nm, and there are a total of 15 resistive heater bridges along the Si waveguide. A higher doping region (P+) in trapezoid shape is connected to the heater bridge on one end, while heavily p-doped (P++) electrodes and vias on the other side.

The integrated BG (IBG) waveguide in our thermal PS design has a waveguide width of $W = 400$ nm, a maximum grating corrugation width of $W_{max} = 800$ nm as sketched in Fig.~\ref{fig:grating2}. The grating period is set at $a = 290$ nm with a duty cycle of 50\%. An apodized super-Gaussian profile is applied to $W_{max}$  to create a chirped IBG for varying strength of slow-light effect. Details of the super Gaussian parameters can be found in the supplemental material. In this work, the concave apodization \cite{jiang2018integrated,chen2022onchip} was chosen which functions to suppress the sideband oscillation at the right-hand side of the band edge. The Bragg grating is situated on a full-height Si with a thickness of 220 nm, while the ridge etching process results in a remaining height of 110 nm for the Si. The half height Si is 2.6 $\mu$m in width centered along the BG.

\section*{Theoretical Analysis of SLTPS in a Nonideal MZI}

\begin{figure}[]
    \centering
    \includegraphics[width=\columnwidth]{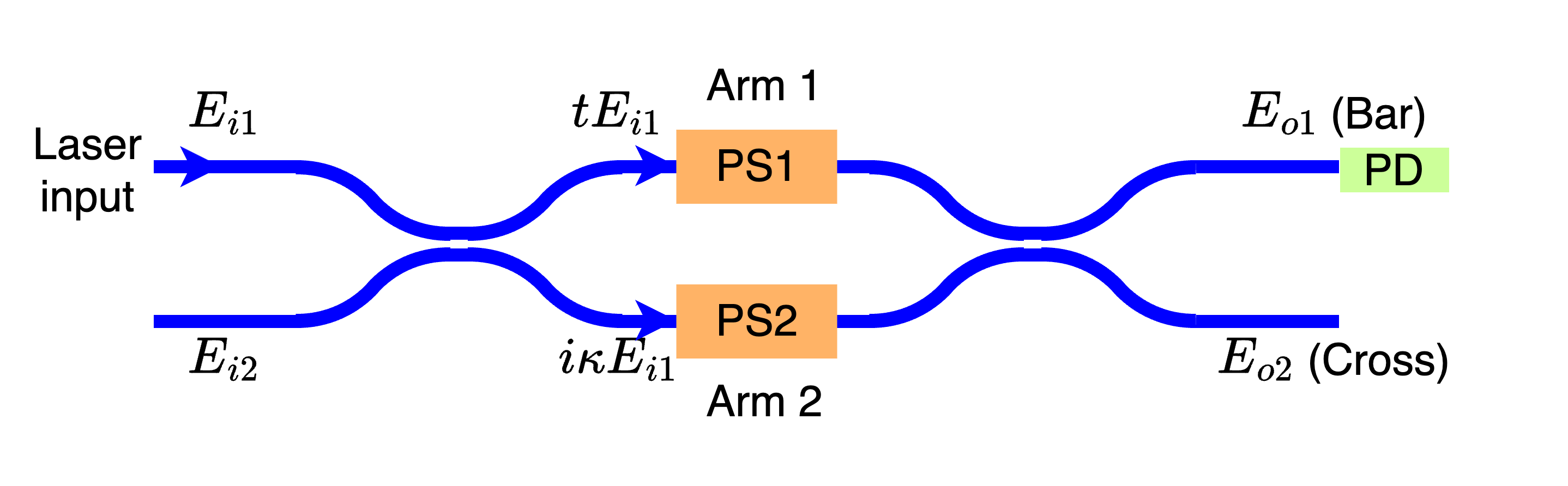}
    \vspace{-10pt}
    \caption{The SLTPS, represented by the boxes denoted PS1 and PS2, within an MZI.}
    \vspace{-10pt}
    \label{fig:mziConfig}
\end{figure}

The terminal characteristics of the SLTPS incorporated in MZI can be calculated from multiplication of the transfer function of each component. A schematic of the SLTPS in MZI is sketched in Fig.~\ref{fig:mziConfig}. It consists of a $2\times2$ optical power splitter at the input and output port and a SLTPS in both MZI arms. The nonidealities of uneven power splitting at input/output ports as well as difference in propagation loss between those two MZI arms are considered in this analysis. 

The electric field at output ports of the MZI, $E_{o1}$ and $E_{o2}$ can be expressed as
\begin{equation}
\resizebox{\columnwidth}{!}{$
    \begin{bmatrix}
        E_{o1}\\
        E_{o2}
    \end{bmatrix} = \begin{bmatrix}
        t & i\kappa\\
        i\kappa & t
    \end{bmatrix}\begin{bmatrix}
        \alpha_1e^{-i\beta_1L_1} & 0\\
        0 & \alpha_2e^{-i\beta_2L_2}
    \end{bmatrix}\begin{bmatrix}
        t & i\kappa\\
        i\kappa & t
    \end{bmatrix}\begin{bmatrix}
        E_{i1}\\
        E_{i2}
    \end{bmatrix}
$}
\label{eqn:mzi}
\end{equation}
where  $\beta_i=2\pi \gamma_in_{\text{eff},i}/\lambda$ is the propagation constant, $\gamma$ is the slowdown factor that arises from slow-light and is a function of wavelength $\lambda$, $n_{\text{eff}}$ is the effective index of the structure, $L$ is the length of the arm, and $\alpha_i$ represents the loss coefficient of arm $i$, defined as $\alpha_i=|E_{\text{out},i}/E_{\text{in},i}|$, where $E_{\text{in},i}$, $E_{\text{out},i}$ are input and output electrical fields of arm $i$ and $i$ can be either 1 or 2. At the optical power splitter, the transmission constant $t$ and the coupling constant $\kappa$ follows the relationship of
\begin{equation}
    t^2+\kappa^2 = 1.
    \label{eqn:constantsRelationship}
\end{equation}

As the heater voltage to the SLTPS varies, there is a concurrent alteration to the PS imparted phase and PS propagation loss due to the stop band edge shift.  With SLTPS in arm 2 under no bias set as the reference, the index and loss change in arm 1 can be evaluated as $\beta_1=\beta_2+\Delta\beta$ and $\alpha_1=1-\Delta\alpha$ while assuming the reference arm is lossless $\alpha_2=1$. As PS length is identical in both arms, we can make the substitution $L_1=L_2=L$. The output of the MZI can be expressed in terms of $\beta_2$, $\Delta\beta$ and $\Delta\alpha$ as

\begin{eqnarray}
    E_{o1}&&=t(tE_{i1}+i\kappa E_{i2})(1-\Delta\alpha)e^{-i(\beta_2+\Delta\beta)L}\nonumber\\
    &&+i\kappa(tE_{i2}+i\kappa E_{i1})e^{-i\beta_2L}
    \label{eqn:arm1}
\end{eqnarray}

\begin{eqnarray}
    E_{o2}&&=t(tE_{i2}\nonumber+i\kappa E_{i1})e^{-i\beta_2L}\\
    &&+i\kappa(tE_{i1}+i\kappa E_{i2})(1-\Delta\alpha)e^{-i(\beta_2+\Delta\beta)L}
    \label{eqn:arm2}
\end{eqnarray}

In our experimental testing, light signal is coupled in from the upper left arm, so we can set $E_{i2}=0$ and $E_{i1}=1$. Eqn.~\ref{eqn:arm1} and~\ref{eqn:arm2} can be simplified as
\begin{figure*}[h!bt]
    \centering
    \begin{subfigure}{0.65\columnwidth}
        \includegraphics[width=\textwidth]{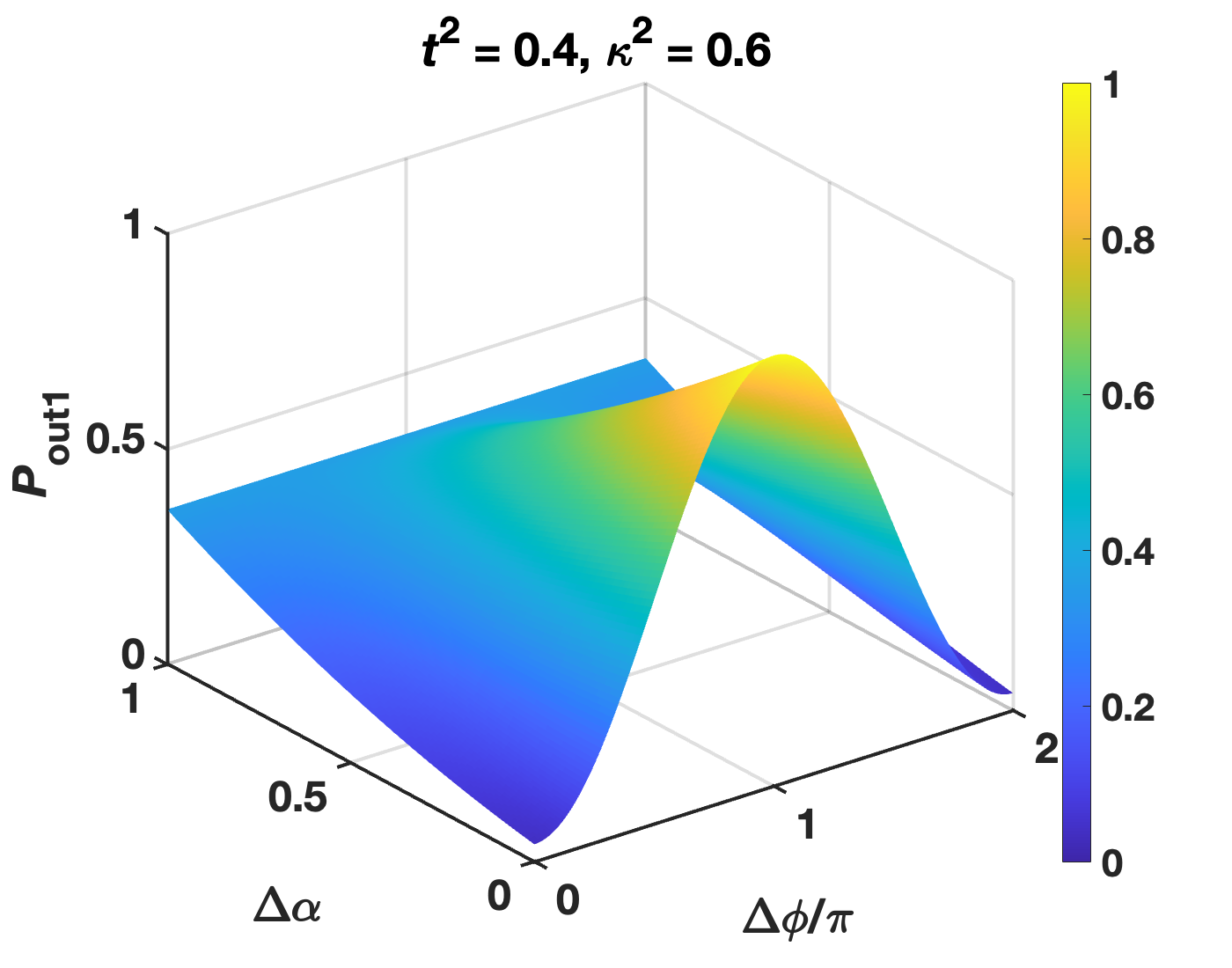}
        \caption{}
        \label{fig:Pout1tsq40Arm1Plot}
    \end{subfigure}
    \begin{subfigure}{0.65\columnwidth}
        \includegraphics[width=\textwidth]{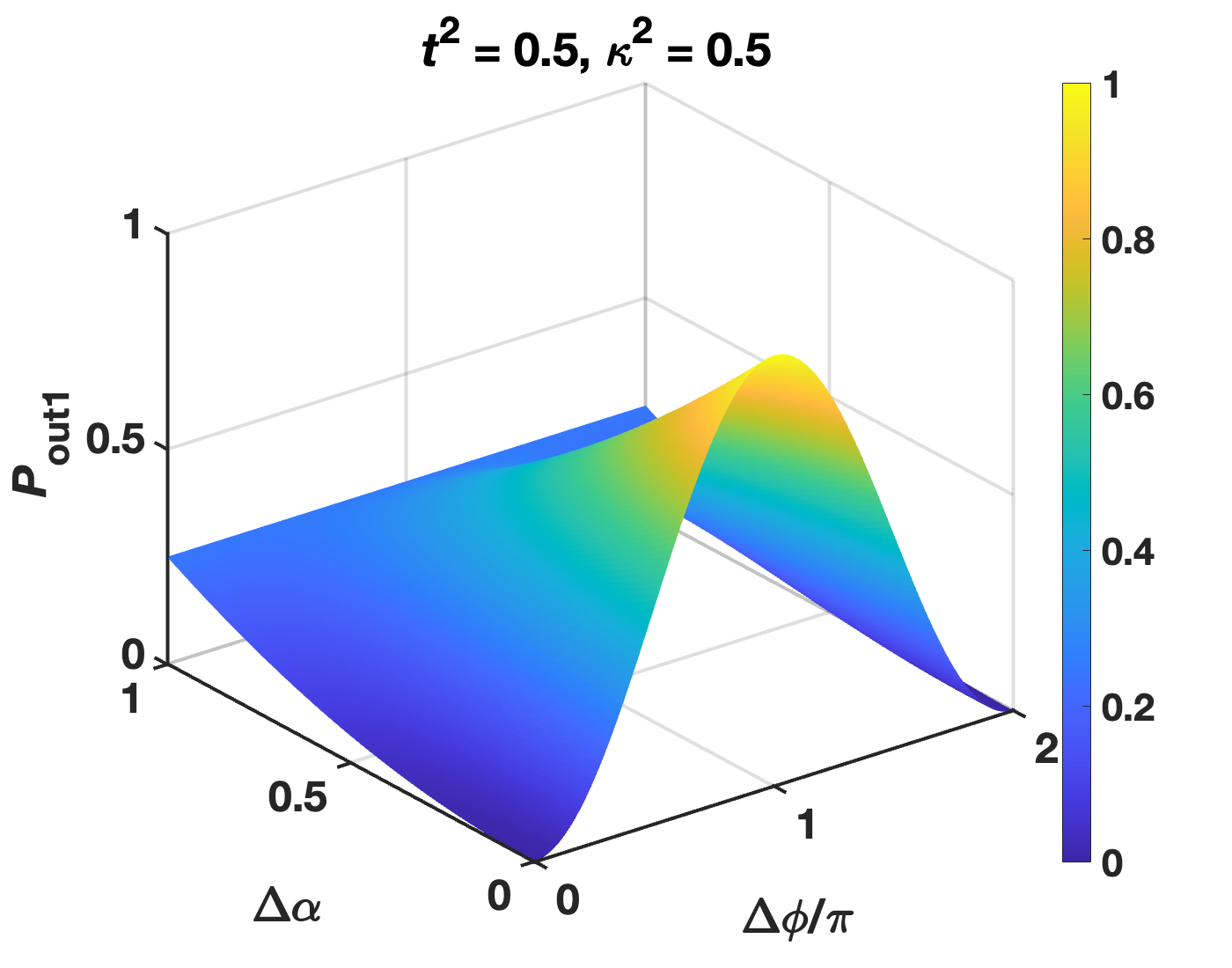}
        \caption{}
        \label{fig:Pout1tsq50Arm1Plot}
    \end{subfigure}
    \begin{subfigure}{0.65\columnwidth}
        \includegraphics[width=\textwidth]{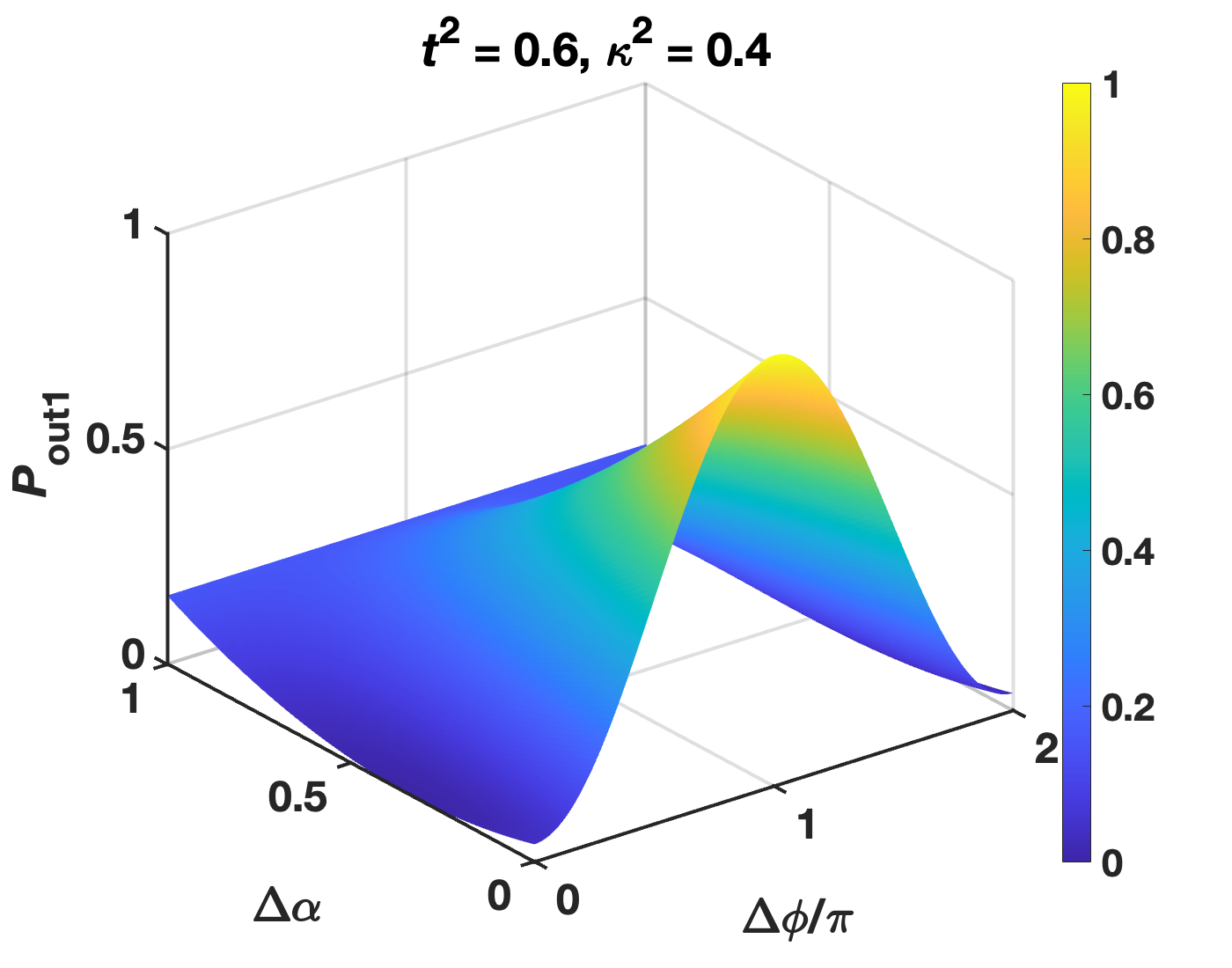}
        \caption{}
        \label{fig:Pout1tsq60Arm1Plot}
    \end{subfigure} \\
    \begin{subfigure}{0.65\columnwidth}
        \includegraphics[width=\textwidth]{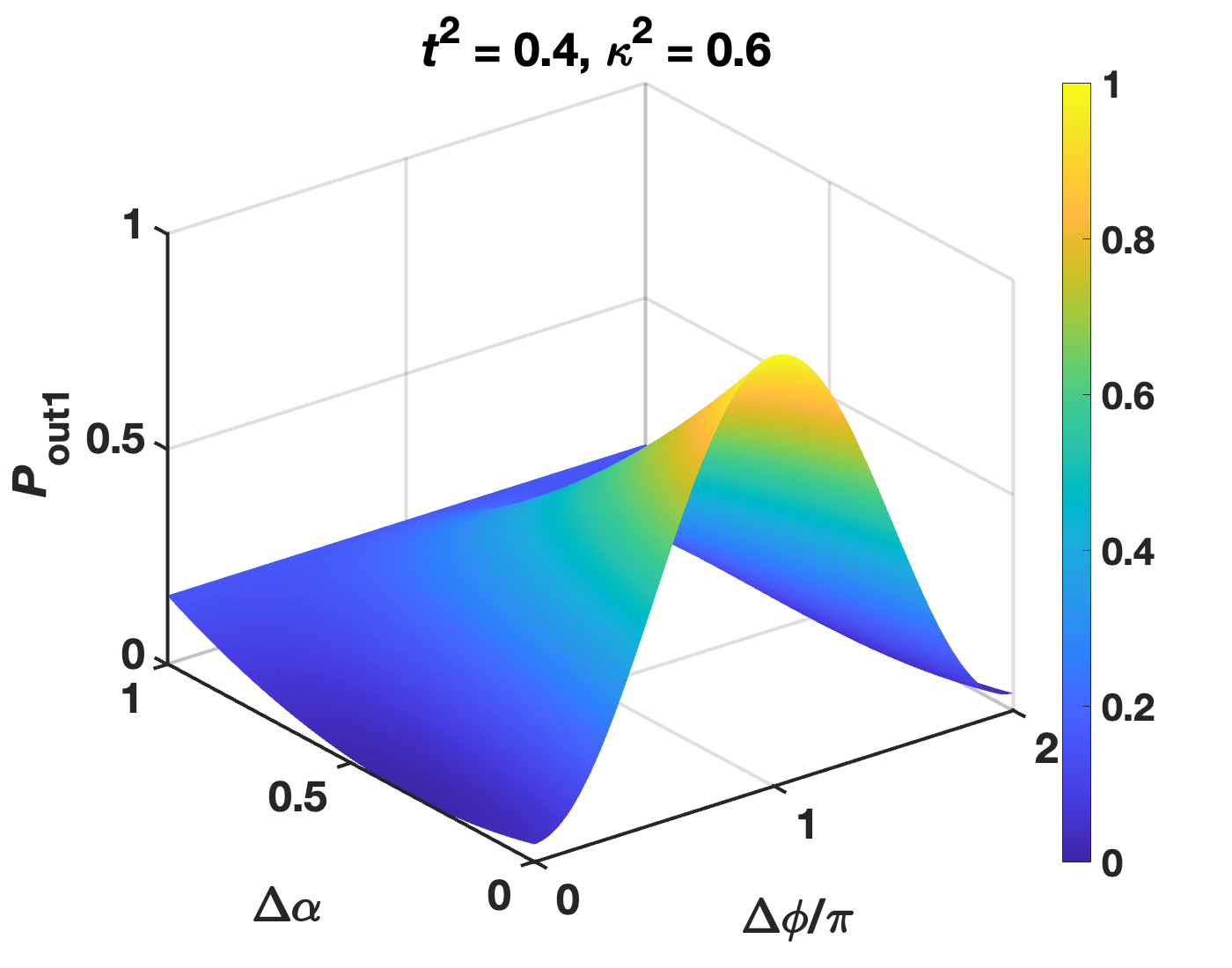}
        \caption{}
        \label{fig:Pout1tsq40Arm2Plot}
    \end{subfigure}
    \begin{subfigure}{0.65\columnwidth}
        \includegraphics[width=\textwidth]{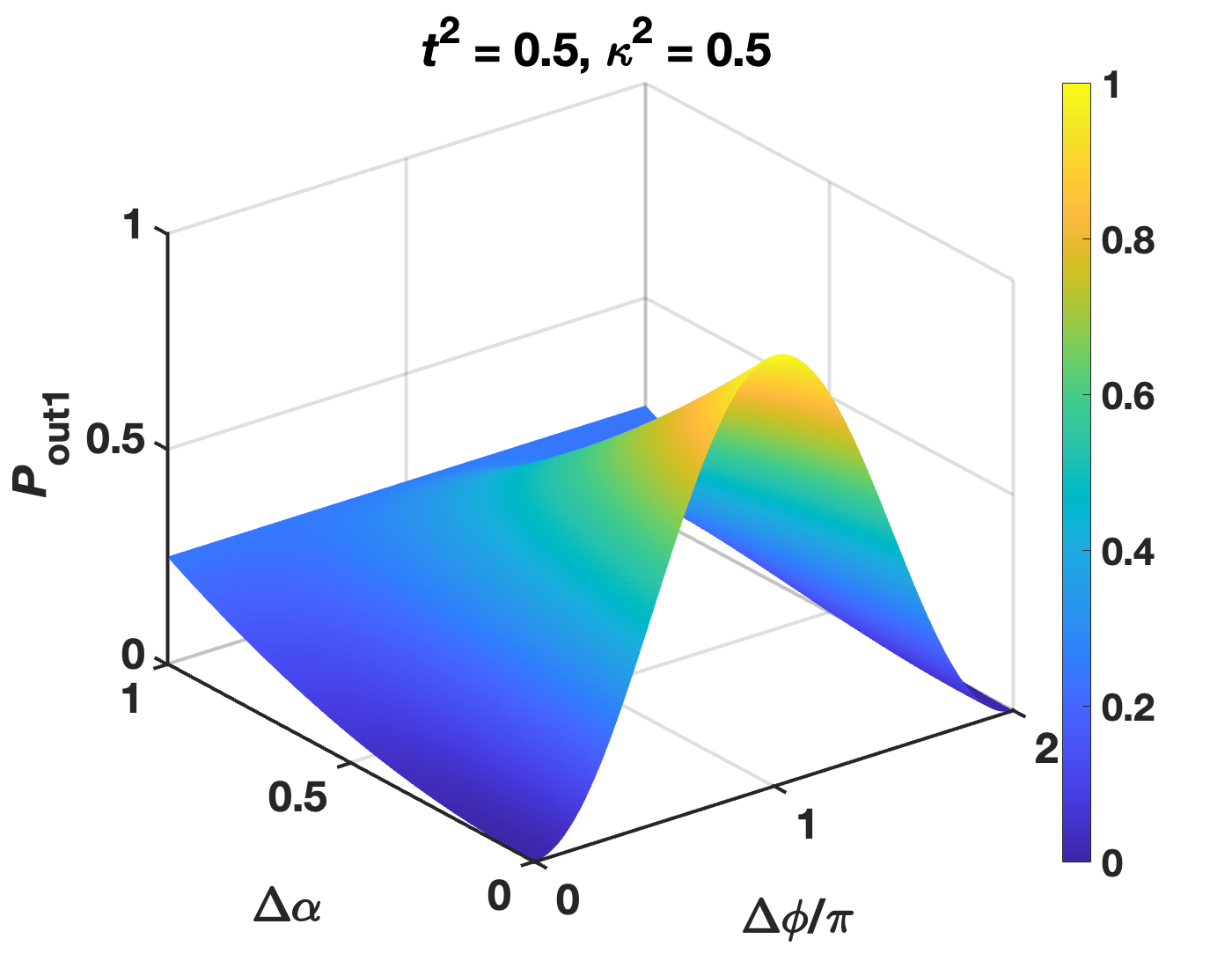}
        \caption{}
        \label{fig:Pout1tsq50Arm2Plot}
    \end{subfigure}
    \begin{subfigure}{0.65\columnwidth}
        \includegraphics[width=\textwidth]{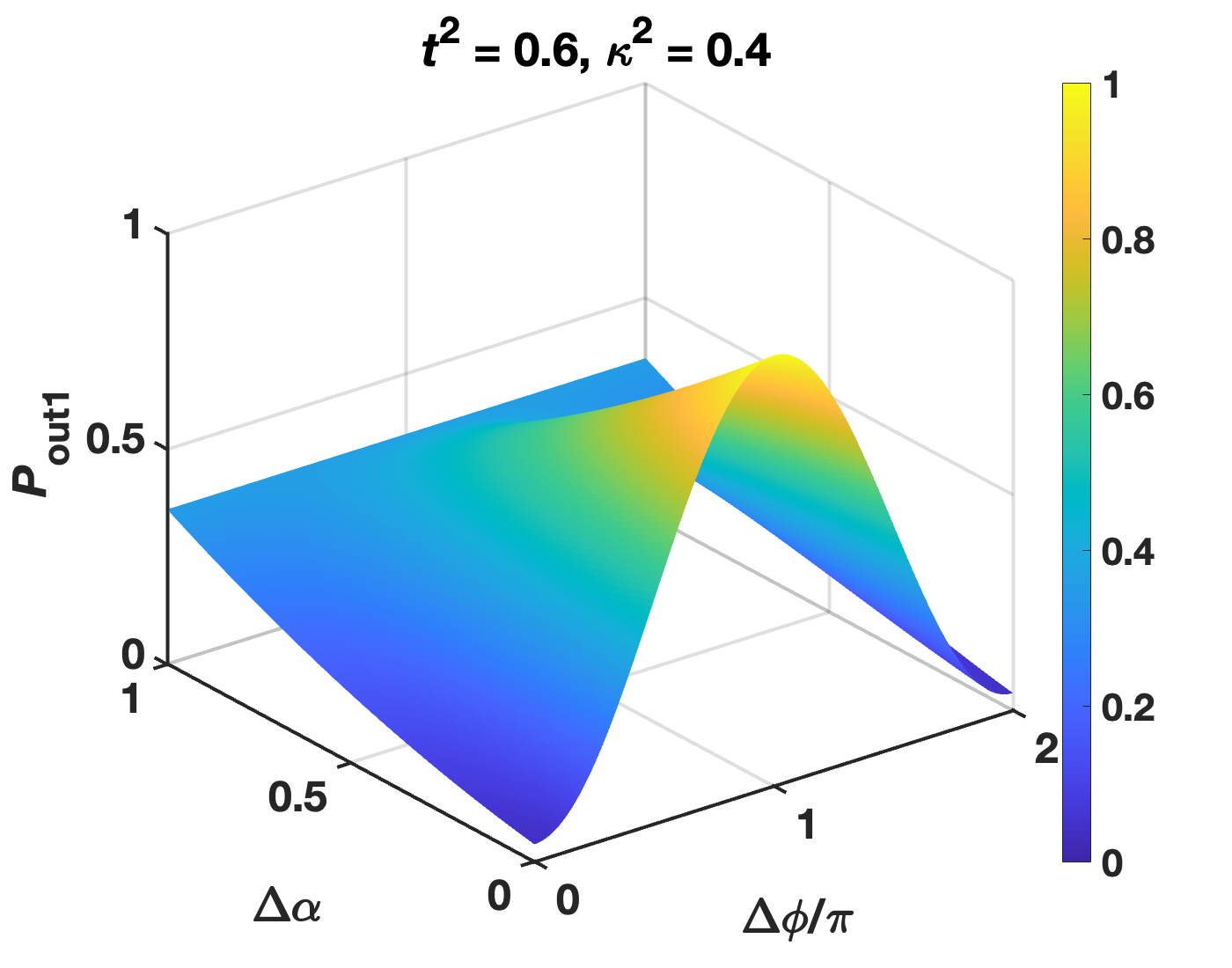}
        \caption{}
        \label{fig:Pout1tsq60Arm2Plot}
    \end{subfigure}
    \caption{$P_{\text{out}1}$ vs. $\Delta\alpha$, $\Delta\phi$ plot when heater bias is applied over the PS in arm 1 with (a) $t^2=0.4$ and $\kappa^2=0.6$, (b)  $t^2=\kappa^2=0.5$ and (c) $t^2=0.6$ and $\kappa^2=0.4$. $P_{\text{out}1}$ vs. $\Delta\alpha$, $\Delta\phi$ when heater bias is applied over the PS in arm 2 with (d) $t^2=0.4$ and $\kappa^2=0.6$, (e)  $t^2=\kappa^2=0.5$ and (f) $t^2=0.6$ and $\kappa^2=0.4$.}
    \label{fig:mziPlotting}
\end{figure*}

\begin{equation}
    E_{o1} = t^2(1-\Delta\alpha)e^{-i(\beta_2+\Delta\beta)L}-(1-t^2)e^{-i\beta_2L}
    \label{eqn:arm1kai}
\end{equation}
\begin{equation}
    E_{o2} = it\sqrt{1-t^2}e^{-i\beta_2L} + it\sqrt{1-t^2}(1-\Delta\alpha)e^{-i(\beta_2+\Delta\beta)L}
    \label{eqn:arm2kai}
\end{equation}
Consider $\Delta\beta L=\Delta\phi$, the MZI output power can be expressed as
\begin{equation}
    P_{\text{out}1}\propto|E_{o1}|^2=|t^2(1-\Delta\alpha)e^{-i\Delta\phi}-(1-t^2)|^2
    \label{eqn:powerArm1}
\end{equation}
\begin{equation}
    P_{\text{out}2}\propto|E_{o2}|^2=t^2(1-t^2)|(1-\Delta\alpha)e^{-i\Delta\phi}+1|^2.
    \label{eqn:powerArm2}
\end{equation}

In this study, the optical power is read out from the bar port, i.e. $P_{\text{out}1}$.  First, assume ideal $2\times2$ optical power splitting, i.e. $t^2=0.5$, $P_{\text{out}1}$ variation will be caused by the difference in loss and thermally induced phase change between the two arm. This condition is plotted in Fig.~\ref{fig:Pout1tsq50Arm1Plot}.  The extra loss in arm 1 due to enhanced slow-light effect would impact the extinction ratio of the MZI optical switch. 

\begin{figure*}[h!bt]
    \centering
    \begin{subfigure}{0.65\columnwidth}
		\centering
		\includegraphics[width=\columnwidth]{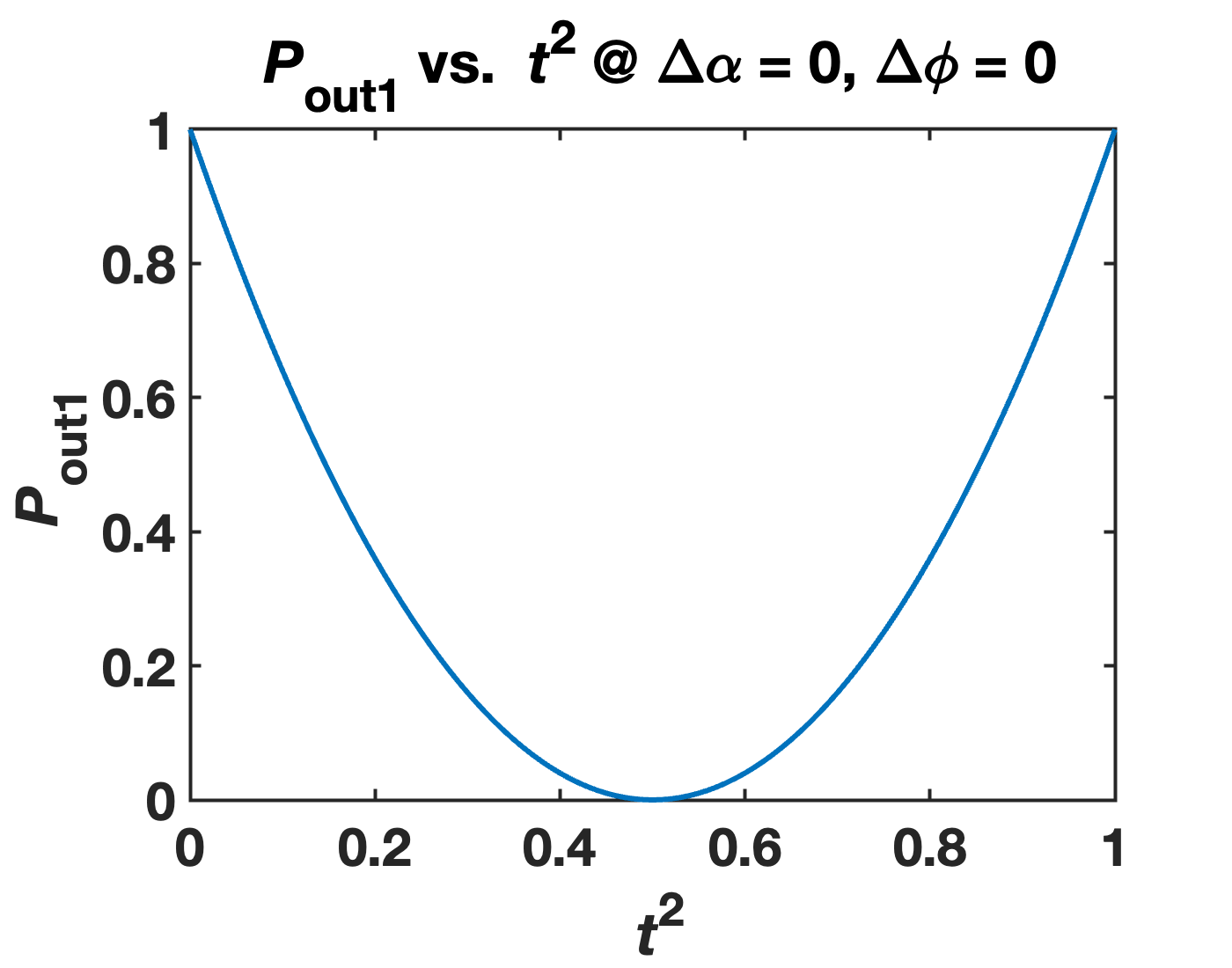}
		\caption{}
		\label{fig:Pout1-tsq}
	\end{subfigure}    
    \begin{subfigure}{0.65\columnwidth}
		\centering
		\includegraphics[width=\columnwidth]{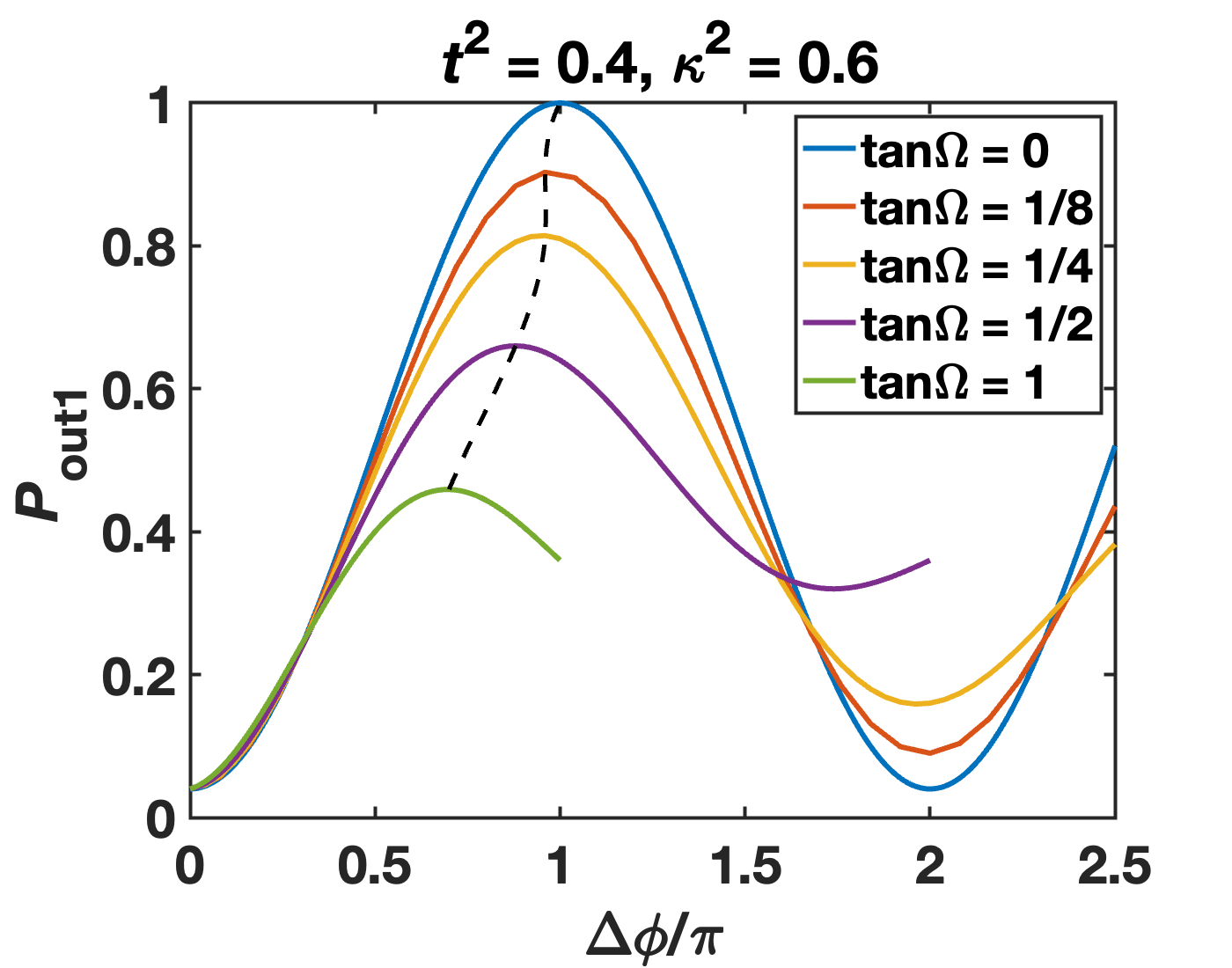}
		\caption{}
		\label{fig:Pout1-phi_tsq40}
	\end{subfigure}
    \begin{subfigure}{0.65\columnwidth}
		\centering
		\includegraphics[width=\columnwidth]{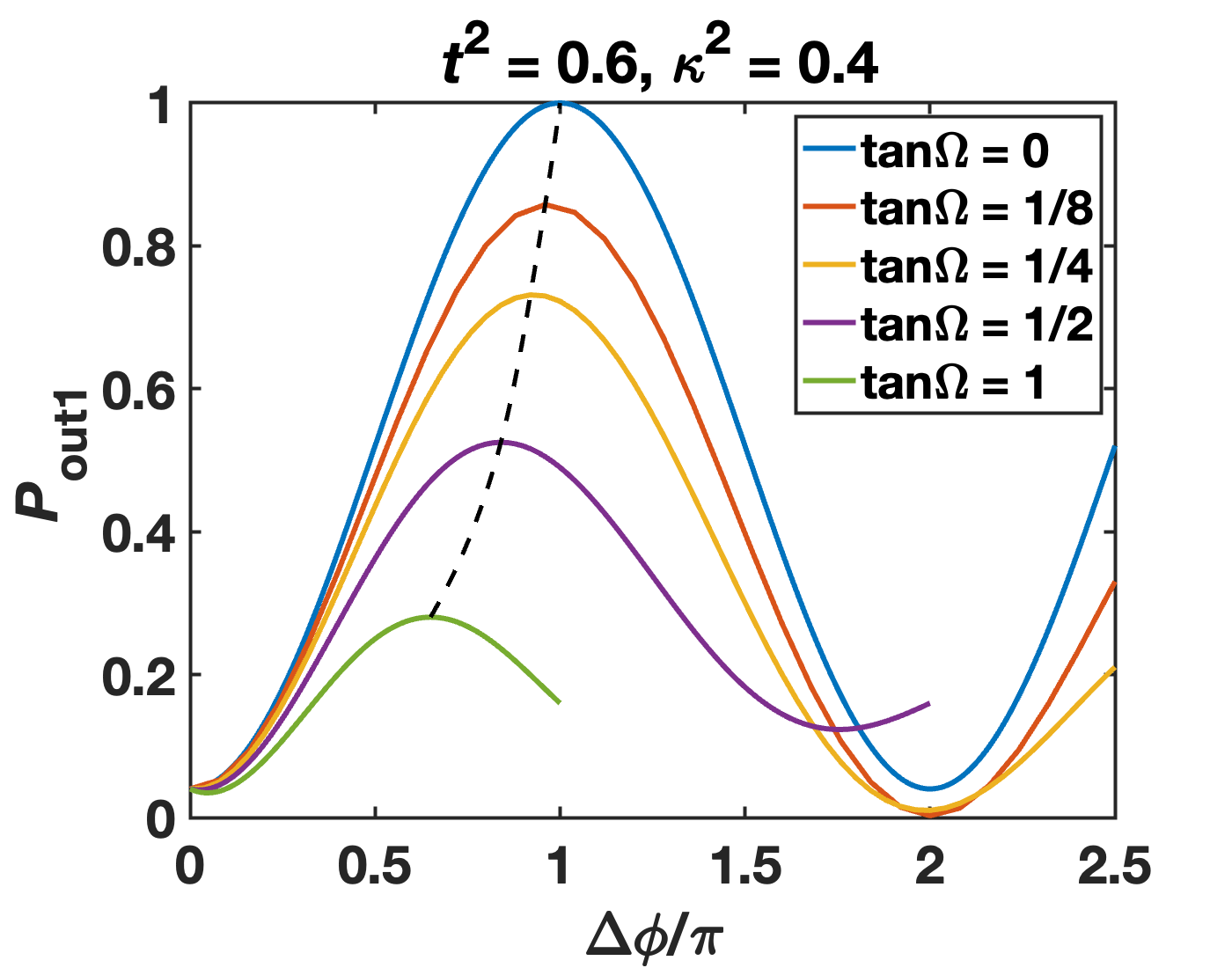}
		\caption{}
		\label{fig:Pout1-phi_tsq60}
	\end{subfigure}\\
    \vspace{-10pt}
    \caption{$P_{\text{out1}}$ plots under different nonideal effects of the MZI. (a) $P_{\text{out1}}$ vs. 
$t^2$ at $\Delta\phi=0, \Delta\alpha=0$. $P_{\text{out1}}$ vs. $\Delta\phi$ plots along different directions on $\Delta\phi-\Delta\alpha$ plane at (b) $t^2=0.4$ and $\kappa^2=0.6$ and (c) $t^2=0.6$ and $\kappa^2=0.4$.}
    \label{fig:Pout1NonidealityPlots}
\end{figure*}

The physical construction of the $2\times2$ optical power splitter is based on directional coupler that exhibits an inevitable deviation from even splitting when not operating at the targeted wavelength by design. Additionally, a number of numerical optimization~\cite{burgwal2017using, perez2020multipurpose} and hardware correction~\cite{bandyopadhyay2021hardware} methods have been studied to address the fabrication induced splitting ratio variation. As examples of uneven splitting, we plot the bar output $P_{\text{out}1}$ for power splitting ratio of $t^2=0.4$ and $\kappa^2 =0.6$ in Fig.~\ref{fig:Pout1tsq40Arm1Plot}, and $t^2=0.6$ and $\kappa^2 =0.4$ in Fig.~\ref{fig:Pout1tsq60Arm1Plot}, respectively. Under no heater bias of arm 1, i.e. $\Delta\alpha=0$ and $\Delta\phi=0$, the plots indicate that $P_{\text{out}1}$ is non zero. $P_{\text{out}1}$ as a function of $t^2$ when $\Delta\alpha=0$ and $\Delta\phi=0$ is shown in Fig.~\ref{fig:Pout1-tsq}.

Consider a fixed $\Delta\alpha$ scenario mathematically, $P_{\text{out}1}$ always increases with $\Delta\phi$ sinusoidally and reaches a local maximum at $\Delta\phi=\pi$. The practical operation of the SLTPS MZI involves simultaneous change of $\Delta\phi$ and $\Delta\alpha$, thus $P_{\text{out}1}$ follows a nonlienar variation in the $\Delta\phi-\Delta\alpha$ space, i.e. a curved plane. We evaluate a set of constant ratios between $\Delta\phi$ and $\Delta\alpha$ at $\text{tan}\Omega=1/8, 1/4, 1/2$ and 1 where $\Omega$ is defined as $\Omega=\frac{\Delta\alpha}{\Delta\phi/\pi}$, as shown in Fig.~\ref{fig:Pout1-phi_tsq40} and~\ref{fig:Pout1-phi_tsq60}.

The $\text{tan}\Omega=0$ case is used as a reference where the maximum of $P_{\text{out}1}$ corresponds to a precise $\pi$ phase shift. We find out for small slope cases, i.e., $\text{tan}\Omega=1/8, 1/4$, $P_{\text{out}1}$ still reaches its maximum near $\Delta\phi=\pi$. However, as the slope approaches 1, the maximum of $P_{\text{out}1}$ significantly shifts to the left, denoted by the dashed black curves in Fig.~\ref{fig:Pout1-phi_tsq40} and~\ref{fig:Pout1-phi_tsq60}, indicating lower than $\pi$ phase shift.
To evaluate whether the maximum of $P_{\text{out}1}$ during experimental testing corresponds to a $\pi$ phase shift, we use the local minimum near $\Delta\phi$ = 2$\pi$ as an indicator. For small $\text{tan}\Omega$ values, such as ${\Delta\phi/\pi}=0, 1/8$ and $1/4$, the local minima of $P_{\text{out}1}$ near $\Delta\phi=2\pi$ are all close to zero. For large $\text{tan}\Omega$ values, not only does the local minima occur at $\Delta\phi$ significantly less than $2\pi$, but also the $P_{\text{out}1}$ value at the local minimum greatly deviates from zero. This can be shown from graph of Fig.~\ref{fig:Pout1-phi_tsq40} and~\ref{fig:Pout1-phi_tsq60}. For our measured SLTPS MZI, we always observed the local minima of $P_{\text{out}1}$ near zero, corresponding to a small slope case. Therefore, in this work, we utilizes the maximum of $P_{\text{out}1}$ to estimate $\pi$ phase shift of the SLTPS.

At a 0 V bias, there is a balanced insertion loss in both arms so $\Delta\alpha=0$. Eqn.~\ref{eqn:powerArm1} and Eqn.~\ref{eqn:powerArm2} can be simplified as  
\begin{equation}
    P_{\text{out}1}\propto|t^2-(1-t^2)|^2
    \label{eqn:powerArm1new}
\end{equation}
\begin{equation}
    P_{\text{out}2}\propto4t^2(1-t^2).
    \label{eqn:powerArm2new}
\end{equation}
For an ideal optical power splitter with even splitting ratio, the output power $P_{\text{out}1}$ = 0 and the cross port power $P_{\text{out}2}$ is 100\%.

Though $P_{\text{out}1}$ is none zero at the bar port due to uneven power splitting, it is of interest to distinguish the $t^2>0.5$ case from the $t^2<0.5$ case so the device design can be adjusted accordingly for future fabrications. This can be accomplished by applying the heater voltage sweep over the PS in arm 2 while keeping arm 1 heater under no bias. Using arm 1 as the reference, $P_{\text{out}1}$ and $P_{\text{out}2}$ with PS2 under heater bias can be written as 
\begin{equation}
    P_{\text{out}1}\propto |(1-t^2)(1-\Delta\alpha)e^{-i\Delta\phi}-t^2|^2
    \label{eqn:powerArm1_bias2}
\end{equation}
\begin{equation}
    P_{\text{out}2}\propto t^2(1-t^2)|(1-\Delta\alpha)e^{-i\Delta\phi}+1|^2.
    \label{eqn:powerArm2_bias2}
\end{equation}

The bar output power $P_{\text{out}1}$ with the PS in arm 2 being biased is plotted for $t^2$ = 0.4, 0.5 and 0.6 in Fig.~\ref{fig:Pout1tsq40Arm2Plot} to~\ref{fig:Pout1tsq60Arm2Plot}, respectively. For the case of $t^2=0.4$ with PS1 biased, $P_{\text{out}1}$ changes monotonically with $\Delta\alpha$ under a constant $\Delta\phi$ (Fig.~\ref{fig:Pout1tsq40Arm1Plot}). $P_{\text{out}1}(\text{min})=0.04$ occurs at $\Delta\alpha=0$, $\Delta\phi=0$. When the heater bias is applied to arm 2, $P_{\text{out}1}$ decreases to zero at $\Delta\alpha=0.33$ and then increases with the increase of $\Delta\alpha$ (Fig.~\ref{fig:Pout1tsq40Arm2Plot}). The zero power of $P_{\text{out}1}$ occurs at which the uneven power splitting is offset by additional optical propagation loss in arm 2. 

\section*{Experimental Results and Discussion}

In this section, the transmission of the slow-light thermal PS at various driving voltages were characterized to quantify the band edge shift with temperature. $V_\pi$ and time constant were measured for different operating wavelengths and their corresponding $P_\pi$ and FOM were calculated.

\subsection*{Thermal response of SLTPS power transmission}
\begin{figure*}[h!bt]
    \centering
     \begin{subfigure}[b]{0.77\columnwidth}
		\centering
		\includegraphics[width=\columnwidth]{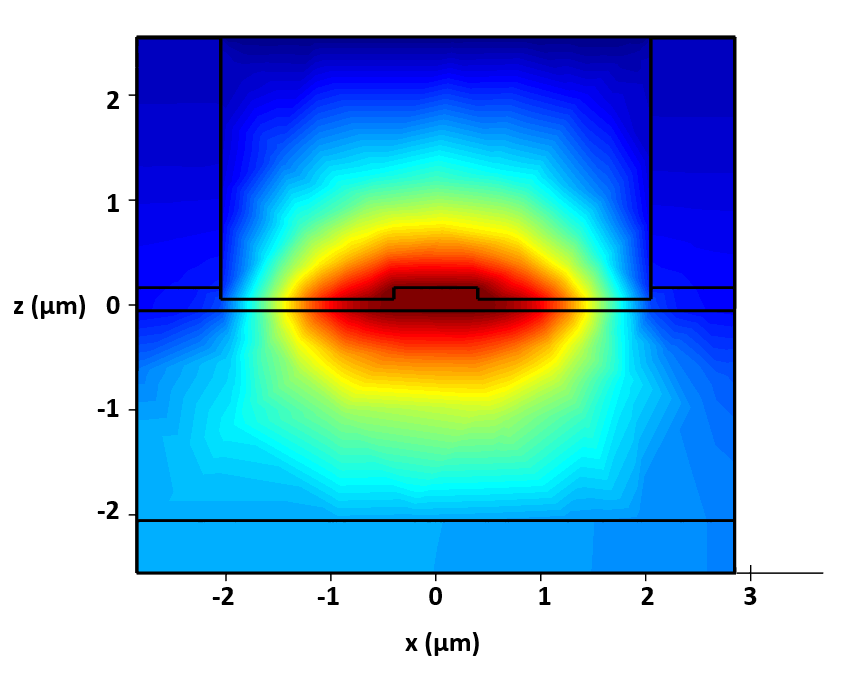}
		\caption{}
		\label{fig:sim2.5}
	\end{subfigure}
     \begin{subfigure}[b]{1.23\columnwidth}
		\centering
		\includegraphics[width=\columnwidth]{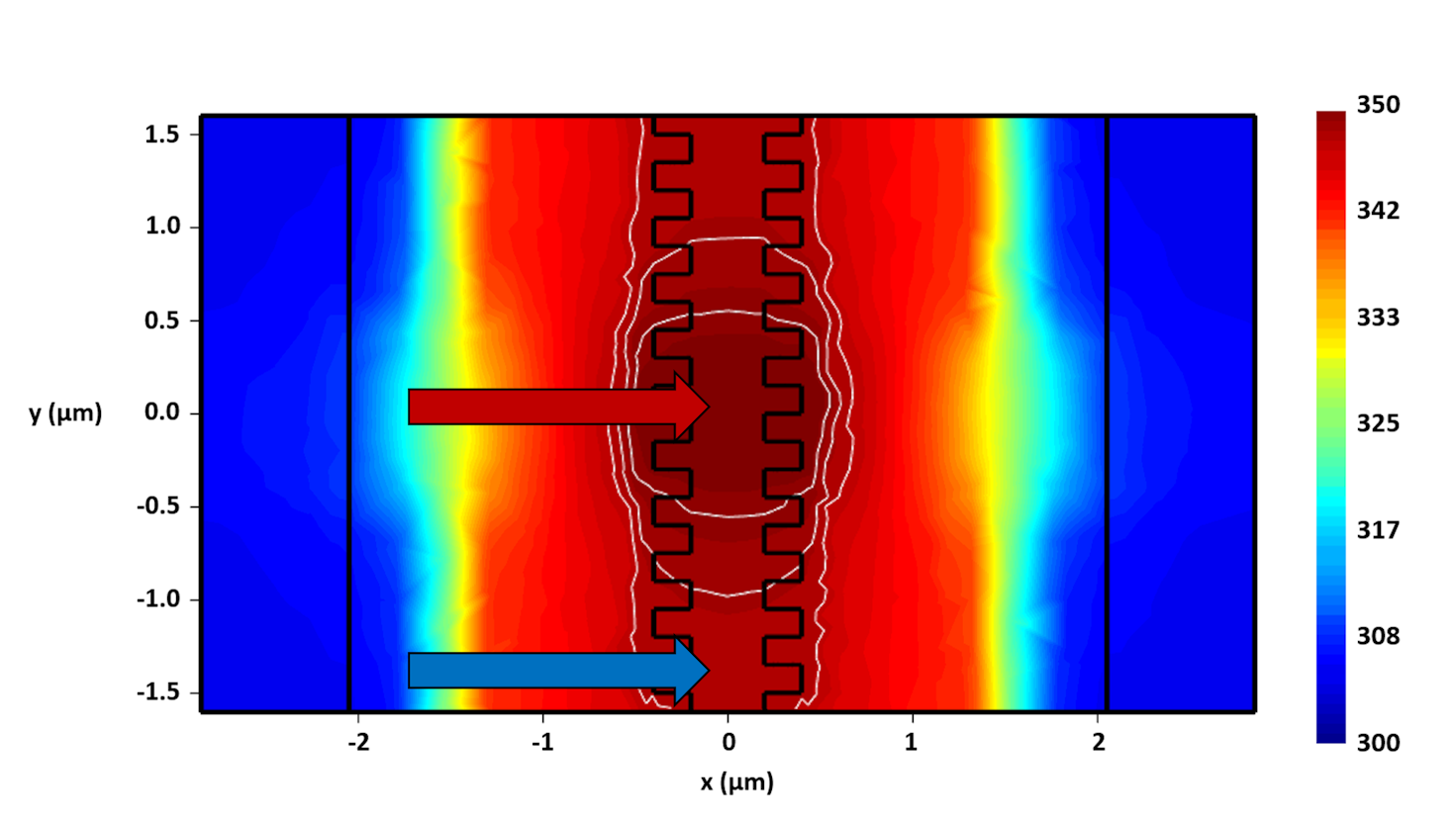}
		\caption{}
		\label{fig:marks}
	\end{subfigure}
    \caption{(a) The cross-sectional view of the thermal profile of the heated BG waveguide arm across the doped Si bridges at an applied voltage of 2.5 V. (b) A top-down view of BG waveguide under heater voltage of 2.5 V. The red arrow points to the region across the heated doped silicon bridge and the blue arrow points out the region in between the doped bridges. The white contour lines correspond to 49$^{\circ}C$, 48$^{\circ}C$, and 47$^{\circ}C$ temperatures. }
    \label{fig:simGraphResults}
\end{figure*}

The thermal modeling of the SLTPS was conducted using the finite element method from Lumerical HEAT\textsuperscript{TM} module where device geometry construction is based on specific design parameters. Details of heat modeling are reported in the supplemental materials. We first investigated the temperature rise under various heater driving conditions, ranging from 0 to 3.5 V. We computed the effective index of the Si waveguide with a width of 400 nm in the absence of grating perturbation across various temperature changes. This serves as a baseline for understanding the change of $V_\pi$ with slow light effect. Using a bias voltage of 2.5 V as an example, the thermal spatial distribution of the SLTPS is shown in Fig.~\ref{fig:sim2.5} and Fig.~\ref{fig:marks}. Joule heating occurs through those discrete heating bridges as the heat source while heat spreads out to the remaining portion of the Si waveguide grating structures via thermal conduction. Since Si is a good heat conductor, we only observed a temperature difference of up to 3$^\circ C$ between the heater bridge and the interstitial regions when the SLTPS reaches temperature exceeding 50$^\circ C$. Although this temperature variation along the Si waveguide may induce a periodic index variation, the periodicity, determined by the bridge interval of 3.2 $\mu$m, has no effect at the operating wavelength of interest.

\begin{figure}[hbt!]
    \captionsetup[subfigure]{justification=centering}
    \centering
    \begin{subfigure}[b]{0.9\columnwidth}
		\centering
		\includegraphics[width=\columnwidth]{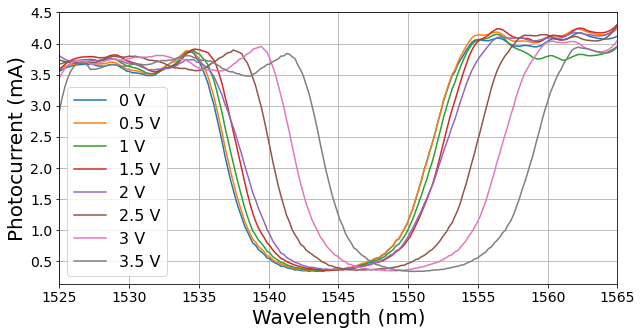}
		\caption{}
		\label{fig:measuredTransmission}
	\end{subfigure}\\
    \begin{subfigure}[b]{0.9\columnwidth}
		\centering
		\includegraphics[width=\columnwidth]{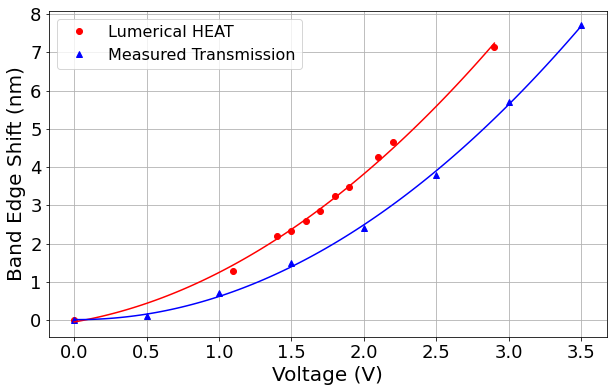}
		\caption{}
		\label{fig:transmissionVersusSim}
	\end{subfigure}\\
    \vspace{-10pt}
    \caption{(a) Transmission spectrum of a single SLTPS at different applied voltages. (b) The change in band edge as a result of the voltage applied.}
    \label{fig:transMeasuredVsSim}
\end{figure}

Under heater bias, the BG transmission spectrum will shift due to the change of material refractive index. The transmission spectrum was studied on a single piece of SLTPS of 50$\mu$m located in near vicinity to the SLTPS-MZI on the same chip to minimize manufacture variation. A red shift of the stop band at elevated temperature was observed and plotted in Fig.~\ref{fig:measuredTransmission}.  Since the apodization is designed to suppress side band oscillation at the right side of the stop band, we only quantify the band edge shift of the right side. The BG waveguide index profile, $n(x,y,z)$, under different bias was first computed using Lumerical HEAT\textsuperscript{TM} and then imported to a waveguide solver, i.e. Lumerical FDTD, to compute the BG spectrum shift. The modeled band edge shift results and the experimentally measured band edge shift are plotted in Fig.~\ref{fig:transmissionVersusSim}.

There is a noticeable deviation in the band edge shift of the measured transmission, indicating that the SLTPS device experiences a smaller temperature increase than predicted by the Lumerical HEAT\textsuperscript{TM} simulation. The plasma etching of the Si BG in device fabrication leaves rounded corners of the grating bars. We modified BG geometry slightly to $W_{min}$ = 410 nm and $W_{max}$ = 780 nm in HEAT simulation to account for plasma etching induced dimension variation and obtained much reduced discrepancy from measured band edge shift. We conclude that the fabrication induced BG geometry variation is the major cause of discrepancy between simulation and experiments. The method we recreated the SLTPS in the simulation could contribute to this issue too. We iteratively adjusted the doping profile of the device until the resistances of the simulated and measured devices matched. Such adjustments can alter properties like thermal conductivity, leading to the band edge shift discrepancy we observe. Nevertheless, the measured band edge shift follows the same trend as the theoretical predication, offering a reliable approach in modifying the device for optimization in the future.

\subsection*{Wavelength selection for operation}

\begin{figure}[]
    \centering
    \includegraphics[width=\columnwidth]{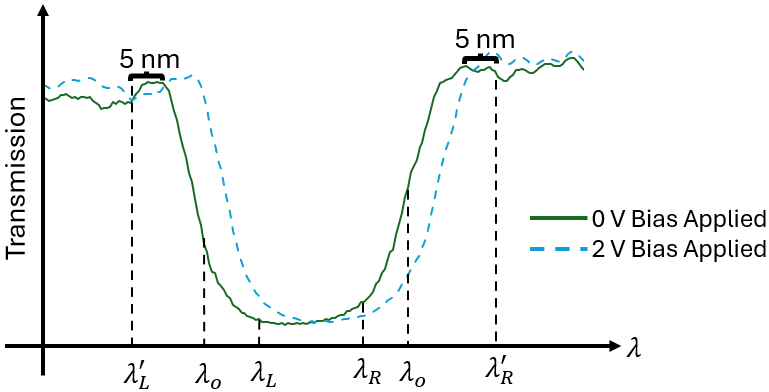}
    \vspace{-10pt}
    \caption{Comparison of wavelengths to use for the left hand side band edge and the right hand side band edge. The turning points of transmission are $\lambda'_L=1534.4$ nm and $\lambda'_R=1553.4$ nm.}
    \vspace{-10pt}
    \label{fig:bandEdgeDifference}
\end{figure}

The thermal PS with BG structure operating in its transmission mode exhibits different responses between the left and right side of the stop band. For conceptual illustration, transmission curves of a single SLTPS piece under no bias and 2 V bias are sketched in Fig.~\ref{fig:bandEdgeDifference}. 
$\lambda_L$ and $\lambda_R$ denotes the left and right side of the stop band edge of the BG, respectively. The slow-light effect is present, albeit diminished, beyond the turning point of the transmission, so we labeled $\lambda'$ at a $\sim$ 5 nm distance away from the edge $\lambda'_R$ and $\lambda'_L$ on Fig.~\ref{fig:bandEdgeDifference} as a spectrum range with sufficiently strong slow-light effect. The operation wavelength $\lambda_o$ of the MZI should lie within the range of $\lambda_R$ < $\lambda_o$ < $\lambda'_R$ of the right band or $\lambda'_L$ < $\lambda_o$ <  $\lambda_L$ of the left band. 

When the side band oscillation is sufficiently suppressed, we can assume the group index $n_g$ is the maximum at $\lambda_R$ and $\lambda_L$ and decreases monotonically when it moves away from the band edge. 
For the right side of the band, when the heater bias to PS1 gradually increases, the slow-light effect at $\lambda_o$ is intensified due to thermally induced red shift, leading to an increased phase shift per unit voltage applied. At a higher $n_g$, the propagation loss in PS1 also increases, thus lowering the $\text{P}_\text{out1}$ at $V_\pi$. At the left band, the increased PS temperature leads to reduced slow-light effect at a fixed operation wavelength $\lambda_o$, but the transmission power of the slow-light PS can increase. Trade-off between $V_\pi$ and PS insertion loss exists for both the right side and left side of band edge. The selection of operation wavelength $\lambda_o$ needs to take this trade-off into consideration. 

\subsection*{$V_\pi$ Measurement and Discussion}
In this work, we only consider a driving mechanism that the heater is biased up to $V_\pi$ while there is no voltage applied on the reference arm. The reference arm can be replaced with a undoped plain waveguide as an alternative design. For the purpose of balancing loss in both arms, a BG with doped bridges is also added to the reference arm. $V_\pi$ can be determined by sweeping the voltage across one PS to identify where it gives the first local maximum of $P_{\text{out}1}$. Under no bias, the stop band edge is identified as $\lambda_L$ = 1537 nm and $\lambda_R$ = 1548 nm from the SLTPS transmission curve (Fig.~\ref{fig:measuredTransmission}). As the BG apodization was designed to reduce side band oscillation in the right side, in this work, we only quantify $V_\pi$ on the right side of the stop band, i.e. $\lambda>\lambda_R$. In our previous study of similar BG structures, the maximum group index $n_g$ near the band edge is measured to $\sim$24  \cite{anderson2022integrated,chen2022onchip} in c-band, representing a slow-light enhancement factor up to 6.

\begin{figure*}[]
    \begin{subfigure}[b]{0.8\columnwidth}
		\centering
		\includegraphics[width=\columnwidth]{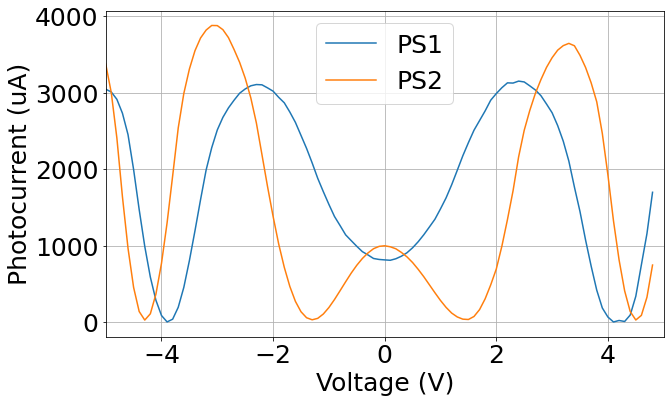}
		\caption{}
		\label{fig:Arm1Arm2comparison}
	\end{subfigure}
 \hspace{60pt}
    \begin{subfigure}[b]{0.8\columnwidth}
		\centering
		\includegraphics[width=\columnwidth]{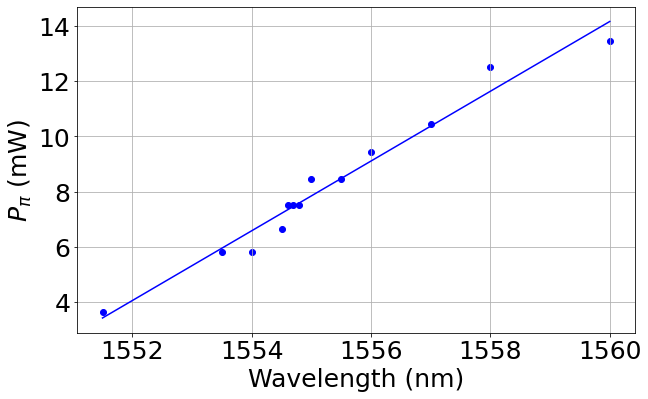}
		\caption{}
		\label{fig:pPiWavelength}
	\end{subfigure}
    \begin{subfigure}[b]{0.9\columnwidth}
		\centering
		\includegraphics[width=\columnwidth]{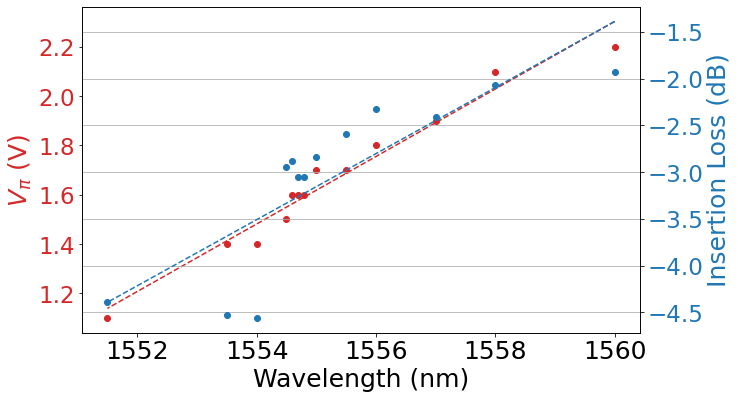}
		\caption{}
		\label{fig:ILTempWave}
	\end{subfigure}
    \caption{(a) Comparison of phase shifter response $P_{\text{out}1}$ as a function of voltage of the SLTPS across arm 1 and arm 2. There is a small difference between those two measurements at V = 0 V due to a slight drift of fiber alignment during test. (b) The relationship between power needed for $\pi$ phase shift and wavelength. (c) A comparison of insertion loss and voltage needed for $\pi$ phase shift at different wavelengths.}
    \vspace{-10pt}
    \label{fig:vpiWavelength}
\end{figure*}

In our testing, a non-zero $P_{\text{out}1}$ was obtained under no bias for all wavelengths, indicating an uneven power splitting in the directional coupler of the MZI as explained by Fig.~\ref{fig:Pout1-tsq}. An example of measured PD output current verse heater bias in arm 1 and arm 2 is plotted in Fig.~\ref{fig:Arm1Arm2comparison}. When bias voltage to PS2 was swept, $P_{\text{out}1}$ goes to zero at $\sim$1.3 V and then increases to a local maximum at $\sim$3.2 V indicating the MZI splitter has $t^2 < 0.5$, following the scenario of Fig.~\ref{fig:Pout1tsq40Arm1Plot} and ~\ref{fig:Pout1tsq40Arm2Plot}. As $V_\pi$ for biasing PS1 is lower than that of biasing PS2, applying heat to PS1 is a preferred operation for better power efficiency.

The stop band edge of the right side is at $\lambda_R$ = 1548 nm while the transmission curve becomes flat at $\lambda$ $\sim$1554 nm. Strong slow-light effect is only expected in a few nm away from the band edge. We first measured $V_\pi$ at $\lambda_o$ = 1580 nm, 25 nm away from the band edge where it has negligible slow-light effect. The tested $V_\pi$ is 2.9 V. This serves as the baseline for the loss comparison in the slow-light region. Next, $V_\pi$ was measured in the spectrum range of 1551.5 nm to 1560 nm where stronger slow-light effect is anticipated. Using PS1 as the working arm while PS2 as the reference arm, the measured $V_\pi$ verse operation wavelength is plotted in Fig.~\ref{fig:ILTempWave}. The measured $V_\pi$ at $\lambda_o$ = 1580 nm is 2.9 V while reduced $V_\pi$ was observed as the operation wavelength approaching the stop band. At $\lambda_o$ = 1551.5 nm, a very low $V_\pi$ = 1.1 V was measured.

The insertion loss of doped Si heater embedded in Si waveguide comes from the scattering loss from the doped bridges and roughness of Si waveguide walls. For SLTPS of this work, the insertion loss increases as $\lambda_o$ is set closer to the stop band. We characterized the SLTPS MZI insertion loss verse wavelength by normalizing the measured transmitted optical power with respect to $\lambda_o$ = 1580 nm and graphed in Fig.~\ref{fig:ILTempWave}. The wavelength dependent power splitting ratio of the MZI also impacts the output power $P_{\text{out}1}$ of the MZI. Relative to transmission at $\lambda_o$ = 1580 nm, the insertion loss at $\lambda$ = 1551.5 nm is 4.4 dB.

The power consumption of the SLTPS heater is calculated using the formula $P_\pi=I_\pi V_\pi$, where $I_\pi$ was measured at $V_\pi$. This accounts for any changes in resistance due to temperature, as the resistivity of the doped heater bridge is inversely proportional to carrier mobility and a change in temperature would lead to mobility variation.

Thermal cross-talk affecting the reference arm can have adverse effects on the phase stability of the MZI and can also lead to increase the optical switching loss.  At a wavelength of 1560 nm and a  $V_\pi$ value of 2.2 V, the temperature increase in PS1 induces thermal cross-talk, measured as 0.315$^\circ$C, into PS2. This corresponds to a thermally induced phase variation of 0.007\% in PS2. For wavelengths below 1555 nm with stronger slow-light effect where $V_\pi$ is below 1.5 V, the temperature in PS1 only increases to 33$^\circ$C above room temperature and does not cause a significant temperature increase in PS2. The decrease in $V_\pi$ due to the slow-light effect is thus beneficial in reducing thermal cross-talk, leading to more compact and densely integrated thermal PS suitable for a variety of applications, and thermal cross-talk can be neglected for this study.

As a comparison, the thermal PS in the AIM PDK library that utilizes similar doped Si heater bridges without slow-light structure has a length of 100 $\mu$m and $V_\pi$ of 3.2 V, giving a power consumption of 30 mW at $V_\pi$. Notably, within the slow-light region, the SLTPS in this work has offered the advantages of half the device length, up to 50\% and 80\% reduction in $V_\pi$ and power consumption, respectively.

\subsection*{Time Constant Measurement}

The figure of merit (FOM), defined as $P_\pi\times\tau$, is commonly utilized to evaluate the efficiency of heaters in a thermally tuned PS. Various assessment criterion have been adopted to ascertain the devices' time constants, including deriving the constant from the 3 dB bandwidth \cite{harris2014efficient}, identifying the time duration at $1/e$  \cite{vancampenhout2010integrated,watts2013adiabatic,masood2013comparison,jacques2019optimization}, or calculating the period in transition between 10\% and 90\% of its final values \cite{sun2010submilliwat,yan2017slow,parra2020atomic}. In this work,  the falling and rising time constants were determined at the $1/e$ and $1-1/e$ points, respectively.

In our study, 20 kHz square wave signals were generated to measure the time constant of the MZI optical switch with SLTPSs. The voltage swing is set from 0 V to $V_\pi$ for each given wavelength. Both rise time $\tau_r$ and fall time $\tau_f$ are measured.The wavelength $\lambda=1580$ nm is again used as a baseline for comparison, with $\tau_r$ and $\tau_f$ measured as 1.49 $\mu$s and 10.62 $\mu$s, respectively. This aligns with the transient time waveform of PS on SOI substrate reported in the literature \cite{jacques2019optimization}. The $\tau_r$ and $\tau_f$ variation with wavelength is plotted in Fig.~\ref{fig:risingFalling}, while the transient response of selected wavelengths is depicted in the supplemental materials. %

\begin{figure}[h!bt]
    \centering
    \includegraphics[width=.9\columnwidth]{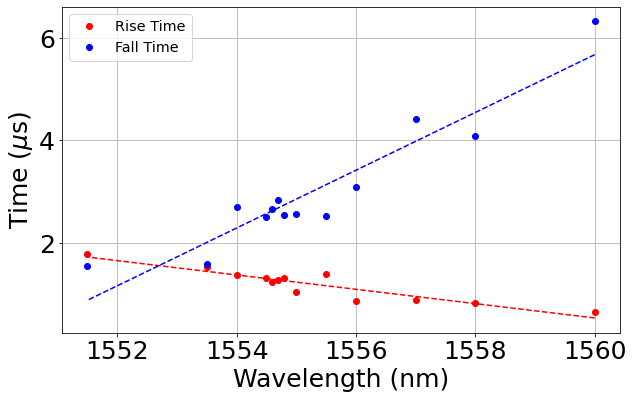}
    \vspace{-10pt}
    \caption{Rising and falling time constants of the SLTPS, highlighting that the falling time constant acts as the limiting factor across all wavelengths except $\lambda = $ 1551.5 nm. A trend line is included to facilitate visualization of this relationship as well as the beneficial effects of slow light on the thermal time constant.}
    \vspace{-10pt}
    \label{fig:risingFalling}
\end{figure}

A notable disparity between the rise time and fall time was observed in our measurements. The rise time, $\tau_r$, for all tested wavelengths consistently falls within a relatively narrow range of 0.8 to 1.8 $\mu$s.  This consistency can be attributed to the embedded doped silicon heating element within the optical waveguide, which constitutes an efficient heating mechanism. The cooling process of the thermal PS involves heat transfer to the SOI substrate and radiation to air via the top silicon layer. We estimate a temperature difference of $\Delta$T of 65.1$^\circ$C needed between those two arms to achieve a $\pi$ phase shift at 1560 nm. A temperature decrease of $\Delta$T of 65.1$^\circ$C  demands a rather prolonged fall time of $\tau_f$ = 6.3 $\mu$s. In contrast, the fall time is reduced to  $\tau_f$ = 1.5 $\mu$s as the estimated $\Delta$T for $\pi$ phase shift is as only as low as 18.4$^\circ$C at an operation wavelength of 1551.5 nm.  

At the atomic level, a pronounced difference exists in photon spectra between the silicon and silicon dioxide  \cite{gunde2000vibrational,henry2008spectral}. Phonons generated in silicon, upon reaching the silicon-silicon dioxide interface of the SOI substrate, encounter a vibrational mode mismatch. This mismatch leads to a considerable portion of phonons being reflected back into the silicon, rather than transmitting through the oxide layer, causing heat trap within the silicon layer, so decelerating the cooling process.  That silicon dioxide's thermal conductivity is significantly lower than that of silicon \cite{yamasue2002thermal,zhu2018thermal} further impedes heat dissipation. The reduced thermal conductivity of silicon dioxide creates a thermal barrier, slowing down the heat dissipation from the structure.

\subsection*{On-Chip Heater Placement Comparison and Discussion}

\begin{table*}[!hbt]
    \caption{Comparison of this study with published literature on thermal PS length, power, time constant, IL and FOM.}\label{tab:comparisonWork2}
    \small
    \begin{tblr}{
  colspec={c X[c] c c c c c},
  vlines,
  hlines,
  vspan=even
}
    \makecell{Location\\of Heater} & Structure & \makecell{Length ($\mu$m)} & \makecell{$P_\pi$ (mW)} & \makecell{Time Constant\\($\mu$s)} & \makecell{IL (dB)} & \makecell{FOM\\(mW$\mu$s)}\\
    \SetCell[r=3]{c}\textbf{In-Plane} & \makecell{NiSi Heater Next to Si \cite{vancampenhout2010integrated}} & 200 & 20 & 2.8 &  0.5 & 56\\
    & \makecell{N++ Si Heater Next to Si \cite{patel2014four}} & 320 & 22.8 & 2.2$^a$ & $<$0.4 & 50.16\\
    & NiCr Heater Next to Si PhC \cite{yan2018efficient} & 10 & 10.05 & 2.4$^b$ & -- & 24.12\\
    \SetCell[r=3]{c}\textbf{Vertical} & Pt Over Suspended Si \cite{sun2010submilliwat} & 100 & 0.54 & 141$^b$ & 2.8 & 76.14\\
    & ITO Over Si \cite{parra2020atomic} & 50 & 9.7 & 5.2$^b$ & $<$0.01 & 50.44\\
    & \makecell{Graphene Heater Over\\Si PhC \cite{yan2017slow}} & 20 & 3.27 & 0.75$^b$ & 5 & 2.37\\
    \SetCell[r=3]{c}\textbf{Embedded} & \makecell{Integrated Heater\\in Adiabatic Si Bend \cite{watts2013adiabatic}} & 10 & 12.7 & 2.4$^a$ & -- & 30.48\\
    & \makecell{Resistive Bridges Across Si \cite{harris2014efficient}} & 61.6 & 24.77 & 2.69$^c$ & 0.23 & 66.63\\
    & This Work & 50 & 3.63 & 1.78$^a$ & 4.4 & 5.1
\end{tblr}
\footnotesize{$^a$Measured using $1/e$, $^b$Measured using 10\% and 90\%, $^c$Measured using 3 dB bandwidth}
\end{table*}

There are several strategies in placing the heater element relative to the Si waveguide which we've categorized into three groups: embedding the heater within the Si waveguide, placing the heater placed in-plane with the Si waveguide, or positioning the heater above the Si waveguide. Comparison of thermal PSs based on different heating materials and physics was reported by \citeauthor{masood2013comparison}\cite{masood2013comparison} Meanwhile, \citeauthor{jacques2019optimization} compared the performance of thermal PSs among several SiPho foundries.\cite{jacques2019optimization} The application of various Si heater placements and their comparison in terms of phase shifter length, $P_\pi$, insertion loss (IL), and figure of merit $P_\pi\times\tau$ are reported in Table~\ref{tab:comparisonWork2}.

Both in-plane and vertical heaters can produce uniform heating as they extend along the length of the Si waveguide. A major design consideration is the separation of the heating element from the Si waveguide region. Incorporating heaters directly within the Si waveguides allows for more efficient and confined thermal tuning, thereby minimizing the device's overall footprint. A trade-off between heating efficiency and optical wave propagation loss is observed in all types of thermal PSs. The use of transparent metals, such as ITO, can help mitigate the optical wave propagation loss to some extend. To date, all three groups of thermal PS designs have been able to achieve a FOM on the order of 50 to 80 mW$\mu$s.

Thermal PS that incorporates the slow-light effect to enhance light-matter interaction have been shown to improve the FOM \cite{yan2017slow,yan2018efficient} with both in-plane and vertical heater placements. This study evaluated thermal phase shifters embedded within a slow-light photonic structure and achieved a remarkable FOM of 5.1 mW$\mu$s, an order of magnitude improvement compared to embedded heater designs without slow-light structures. Though the measured $n_g$ suggests a slow-down factor of 3$\sim$6 \cite{chen2022onchip,anderson2022integrated}, the power consumption,  following quadratically change with $V_\pi$, can be reduced significantly due to the slow-light effect.

The comparisons in Table~\ref{tab:comparisonWork2} highlight the efficiency of the SLTPS within the slow-light region, which lowers $\Delta T_\pi$, the time constant in transient operation, and $P_\pi\times\tau$. However, as the operating wavelength approaches the band edge to maximize the slow-light effect, the transmission loss of the SLTPS increases, illustrating a fundamental trade-off in this design. Based on system requirements for power consumption, footprint, and signal-to-noise ratio, the optimal operating wavelength needs to be carefully chosen.

\section*{Conclusion}
In this study, we performed a thorough study of SLTPSs in MZIs including the non-idealities of uneven optical power splitting, unbalanced propagation loss in both arms and wavelength dependent loss. Through thermal modeling, we simulated the red shift at the photonic band edge, compared it with experimental results and emphasized the critical role of BG geometry in band gap shift and temperature rise. Operation in the left-side verse the right-side of the stop band is compared conceptually with a focus on the slow-light effect on the right-side band edge for experimental demonstration. The trade-off between $V_\pi$ and MZI insertion loss was analyzed. The $V_\pi$ measurement demonstrated a very low voltage of 1.1V at $\lambda = 1551.5$ nm with an insertion loss of 4.4 dB and a power consumption of $P_\pi$ = 3.63 mW. Operation at this wavelength gives a FOM of 5.1 mW$\mu$s. We also quantified the dynamic performance of the SLTPS-MZI, with rising and falling time constants measured at 1.78 $\mu$s and 1.55 $\mu$s, respectively. This work presents the first study of a MZI matrix/switch using slow-light effect of Bragg grating for $V_\pi$ reduction while fully utilizing commercial foundry for device fabrication. The device geometry and operation condition can be further optimized to reduce loss and improve $V_\pi$ performance. Error connection designs developed in non-slow light based MZI can also be applied to the SLTPS-MZI in future work.

\begin{acknowledgement}

The authors thank AIM Photonics for providing space on the Constanza MPW run, SMART scholarship to support graduate students research for Alexander Chen and Nick Gangi, and partial funding support from SUNY Research Foundation and New York Focus Center. We also like thank former group member Dr. Stephen Anderson for many insightful discussions. 

\end{acknowledgement}

\begin{suppinfo}

Details on the super-Gaussian apodization profile and its parameters; simulation details of the resistive heater design using Lumerical HEAT\textsuperscript{TM} software; procedure and results of the time constant measurements using a 20 kHz square wave signal.

\end{suppinfo}

\bibliography{achemso-demo}

\end{document}